\documentclass[sigconf]{acmart}

\usepackage[dvipsnames]{xcolor}
\usepackage{graphicx}     
\usepackage{subcaption}   
\usepackage{mwe} 
\usepackage{enumitem} 
\usepackage{mdframed}
\usepackage{multirow}
\usepackage{tabularx}
\usepackage{adjustbox}
\usepackage{tcolorbox}
\usepackage{hyperref}
\usepackage{wrapfig}

\newcommand{\distance}{5pt}
\newcommand{\summary}[1]{
 \begin{center}
  \begin{tcolorbox}[colback=gray!10,colframe=black!25,width=1\columnwidth,arc=1mm, auto outer arc,boxrule=0.5pt,boxsep=2pt,left=3pt,right=3pt,top=0pt,bottom=0pt]
  \textbf{SUMMARY:} #1
  \end{tcolorbox}
 \end{center}
}

\usepackage{bm}
\definecolor{myblue}{RGB}{255,255,255}
\definecolor{myyellow}{HTML}{FFF2CC}

\newcommand{\inlinecode}[1]{\texttt{\small #1}\xspace}

\newlength{\ColorBoxDepthReference}
\newlength{\ColorBoxHeightReference}
\newlength{\Width}%
\newcommand{\MyColorBox}[2][red]%
{%
	\settowidth{\Width}{#2}%
	\colorbox{#1}%
	{%
		\raisebox{-\ColorBoxDepthReference}%
		{%
			\parbox[b][\ColorBoxHeightReference+\ColorBoxDepthReference][c]{\Width}{\centering#2}%
		}%
	}%
}
\newcommand{\tool}{\textsc{RepoLens}\xspace}

\AtBeginDocument{%
  }

\setcopyright{acmlicensed}
\copyrightyear{2018}
\acmYear{2018}
\acmDOI{XXXXXXX.XXXXXXX}
\acmConference[Conference acronym 'XX]{Make sure to enter the correct
  conference title from your rights confirmation email}{June 03--05,
  2018}{Woodstock, NY}
\acmISBN{978-1-4503-XXXX-X/2018/06}

\begin{document}

\title{Extracting Conceptual Knowledge to Locate Software Issues}

\author{Ying Wang}
\email{yingwang25@m.fudan.edu.cn}
\affiliation{%
  \institution{Fudan University}
  \country{China}
}

\author{Wenjun Mao}
\email{wjmao25@m.fudan.edu.cn}
\affiliation{%
  \institution{Fudan University}
  \country{China}
}

\author{Chong Wang}
\email{chong.wang@ntu.edu.sg}
\affiliation{%
  \institution{Nanyang Technological University}
  \country{Singapore}
}

\author{Zhenhao Zhou}
\email{zhzhou24@m.fudan.edu.cn}
\affiliation{%
  \institution{Fudan University}
  \country{China}
}

\author{Yicheng Zhou}
\email{yichengzhou23@m.fudan.edu.cn}
\affiliation{%
  \institution{Fudan University}
  \country{China}
}

\author{Wenyun Zhao}
\email{wyzhao@fudan.edu.cn}
\affiliation{%
  \institution{Fudan University}
  \country{China}
}

\author{Yiling Lou}
\email{yilinglou@fudan.edu.cn}
\affiliation{%
  \institution{Fudan University}
  \country{China}
}

\author{Xin Peng}
\email{pengxin@fudan.edu.cn}
\affiliation{%
  \institution{Fudan University}
  \country{China}
}

\begin{abstract}
Issue localization, which identifies faulty code elements such as files or functions, is critical for effective bug fixing. While recent LLM-based and LLM-agent-based approaches improve accuracy, they struggle in large-scale repositories due to \textit{concern tangling}, where relevant logic is buried in large functions, and \textit{concern scattering}, where related logic is dispersed across files.

To address these challenges, we propose \tool, a novel approach that abstracts and leverages conceptual knowledge from code repositories. \tool decomposes fine-grained functionalities and recomposes them into \textit{high-level concerns}, semantically coherent clusters of functionalities that guide LLMs. It operates in two stages: an offline stage that extracts and enriches conceptual knowledge into a repository-wide knowledge base, and an online stage that retrieves issue-specific terms, clusters and ranks concerns by relevance, and integrates them into localization workflows via minimally intrusive prompt enhancements.
We evaluate \tool on SWE-Lancer-Loc, a benchmark of 216 tasks derived from SWE-Lancer. \tool consistently improves three state-of-the-art tools, namely AgentLess, OpenHands, and mini-SWE-agent, achieving average gains of over 22\% in Hit@k and 46\% in Recall@k for file- and function-level localization. It generalizes across models (\textit{GPT-4o}, \textit{GPT-4o-mini}, \textit{GPT-4.1}) with Hit@1 and Recall@10 gains up to 504\% and 376\%, respectively. Ablation studies and manual evaluation confirm the effectiveness and reliability of the constructed concerns.
\end{abstract}




\maketitle

\section{Introduction}\label{sec-intro}


Issue localization, which involves identifying the faulty code elements (e.g., files or functions) in software, is a pivotal task in software quality assurance~\cite{DBLP:journals/tse/WongGLAW16, 10114452}. Given a codebase and a issue report (e.g., a user-reported symptom description), automated issue localization techniques generate a ranked list of potentially faulty code elements, ordered by their likelihood of being buggy. Accurate issue localization is especially critical for effective bug fixing, as resolving a bug necessitates the precise identification of its corresponding code location. 

Recently, with advancements in large language models (LLMs), LLM-based and LLM-agent-based approaches~\cite{DBLP:journals/corr/abs-2407-01489, DBLP:conf/iclr/0001LSXTZPSLSTL25, DBLP:journals/corr/abs-2405-15793, DBLP:journals/corr/abs-2409-02977} have shown significant promise in improving the accuracy of issue localization, as evidenced by their performance on SWE-Bench~\cite{DBLP:conf/iclr/JimenezYWYPPN24}. However, achieving accurate localization remains highly challenging in large-scale \textit{application} software due to inherent design complexity. The recent SWE-Lancer benchmark~\cite{DBLP:journals/corr/abs-2502-12115} shows that even frontier models still struggle to solve complex tasks in real-world repositories (e.g., the Expensify project~\cite{expensify} with 2.04 million lines of code). A primary difficulty lies in capturing issue-related concerns that span multiple components or files, which is essential for understanding the root cause~\cite{DBLP:journals/corr/abs-2502-12115}.

More broadly, the challenges of issue localization and the limitations of existing approaches can be summarized as follows. In large-scale code repositories, localization is hindered by the lack of a high-level conceptual view and by two core difficulties: concern tangling and concern scattering. Concern tangling arises when critical logic is buried within large, multi-purpose functions, making it difficult to isolate the relevant code. Concern scattering occurs when semantically related logic is dispersed across multiple files or modules, forcing developers to piece together fragmented functionalities to identify the root cause. A motivating example from the Expensify repository demonstrates that these challenges not only slow down human developers but also restrict the effectiveness of LLM-based approaches such as mini-SWE-agent~\cite{DBLP:journals/corr/abs-2405-15793}, which frequently mislocalize issues due to inefficient and unguided codebase exploration (see Section~\ref{sec-motiv}).

\textbf{\textit{Key Idea.}}
To address these challenges and limitations, we propose abstracting conceptual knowledge by decomposing fine-grained functionalities and recomposing them into high-level concerns. A concern can generally be categorized into two types: (i) a set of interdependent functionalities responsible for a specific feature, or (ii) a set of crosscutting functionalities that address certain aspects of the codebase (e.g., logging in aspect-oriented programming). The complete definition of a concern is provided below:
\begin{mdframed}
[linecolor=myblue!50,linewidth=2pt,roundcorner=10pt,backgroundcolor=gray!20,leftmargin=0pt,rightmargin=0pt,innertopmargin=2pt,innerbottommargin=2pt,innerleftmargin=2pt,innerrightmargin=2pt]
\small
\hypertarget{concern-def}{\textsc{\textbf{Concern Definition:}}} 

\noindent A concern is a cohesive set of functionalities in code that are conceptually related to a specific term (entity, object, or domain concept).
\begin{itemize}[label=-,leftmargin=10pt]
    \item These functionalities may be scattered across different parts of the codebase, but together they are concerned with a single responsibility or area of interest.
    \item A Concern can be composed of atomic operations and low-level details, which, when aggregated, form a higher-level functional semantics or business logic tied to that term. 
\end{itemize}
\end{mdframed}
During issue localization, the most relevant concerns guide exploration by providing a high-level conceptual view, thereby improving both efficiency and accuracy.

\textbf{\textit{Approach.}} 
Based the idea, we propose \tool, a novel approach to enhance issue localization by abstracting and leveraging conceptual knowledge from code repositories. Rather than treating the repository as a flat collection of files and functions, \tool constructs high-level representations called concerns, which are semantically coherent clusters of functionalities that provide explicit guidance for LLMs and LLM-based agents. \tool operates in two stages: an offline stage that extracts and enriches conceptual knowledge by decoupling fine-grained functionalities associated with conceptual terms and building a repository-wide knowledge base with term explanations, and an online stage that performs four steps: (i) issue-specific term retrieval from the knowledge base, (ii) conceptual concern clustering, (iii) concern ranking by relevance to the issue, and (iv) concern-enhanced issue localization via minimally intrusive prompt integration.

\textbf{\textit{Evaluation.}}
We conduct extensive experiments to evaluate the effectiveness of \tool in enhancing issue localization. To this end, we first construct a benchmark, SWE-Lancer-Loc, derived from SWE-Lancer~\cite{DBLP:journals/corr/abs-2502-12115} by applying a set of filtering rules. This process yields 216 issue localization tasks, each containing an average of 1.42 gold files and 1.55 gold functions. Experimental results show that, using \textit{GPT-4o} as the base model, \tool consistently improves localization performance across three state-of-the-art tools—AgentLess~\cite{DBLP:journals/corr/abs-2407-01489}, OpenHands~\cite{DBLP:conf/iclr/0001LSXTZPSLSTL25}, and mini-SWE-agent~\cite{DBLP:journals/corr/abs-2405-15793}—achieving average relative gains of over 22\% in Hit@k and 46\% in Recall@k for file-level and function-level localization across $k$ values from 1 to 10 (\textbf{RQ1}).
Model generalizability experiments further confirm that \tool consistently enhances issue localization across all three base models (\textit{GPT-4o}, \textit{GPT-4o-mini}, and \textit{GPT-4.1}) and the three tools, with relative gains ranging from 2.76\% to 504.35\% for Hit@1 and 2.83\% to 376.13\% for Recall@10 across file- and function-level tasks (\textbf{RQ2}).
An ablation study comparing concerns extracted by \tool with two variants of functionality summaries demonstrates the contribution of conceptual term explanation and conceptual concern clustering (\textbf{RQ3}). Finally, manual annotation of the constructed concerns confirms their reliability in terms of correctness, completeness, and conciseness (\textbf{RQ4}).

In summary, this paper makes the following main contributions:
\begin{itemize}[leftmargin=15pt, topsep=2pt, itemsep=2pt]
    \item \tool, a novel approach to enhance issue localization by abstracting and leveraging conceptual knowledge from code repositories. It decomposes fine-grained functionalities in the codebase and recomposes them into high-level concerns for a given issue, which can be integrated into various LLM-driven workflows or agent-based localization tools.  
    \item Extensive experimental results demonstrating that the constructed concerns consistently improve the performance of three recent localization approaches across three different base models for both file- and function-level localization. With \tool, mini-SWE-agent achieves a new state-of-the-art Hit@1 of 41.67\% for file-level and 25.93\% for function-level localization using \textit{GPT-4.1}, corresponding to relative improvements of 57.90\% and 30.24\%, respectively.  
    \item A dataset of generated term-centric functionalities and clustered conceptual concerns, serving as a resource for future related research using the SWE-Lancer benchmark.
\end{itemize}

\section{Motivation}\label{sec-motiv}

\begin{figure}
    \centering
    \footnotesize
    \begin{subfigure}[b]{0.49\textwidth} 
        \includegraphics[width=1\linewidth]{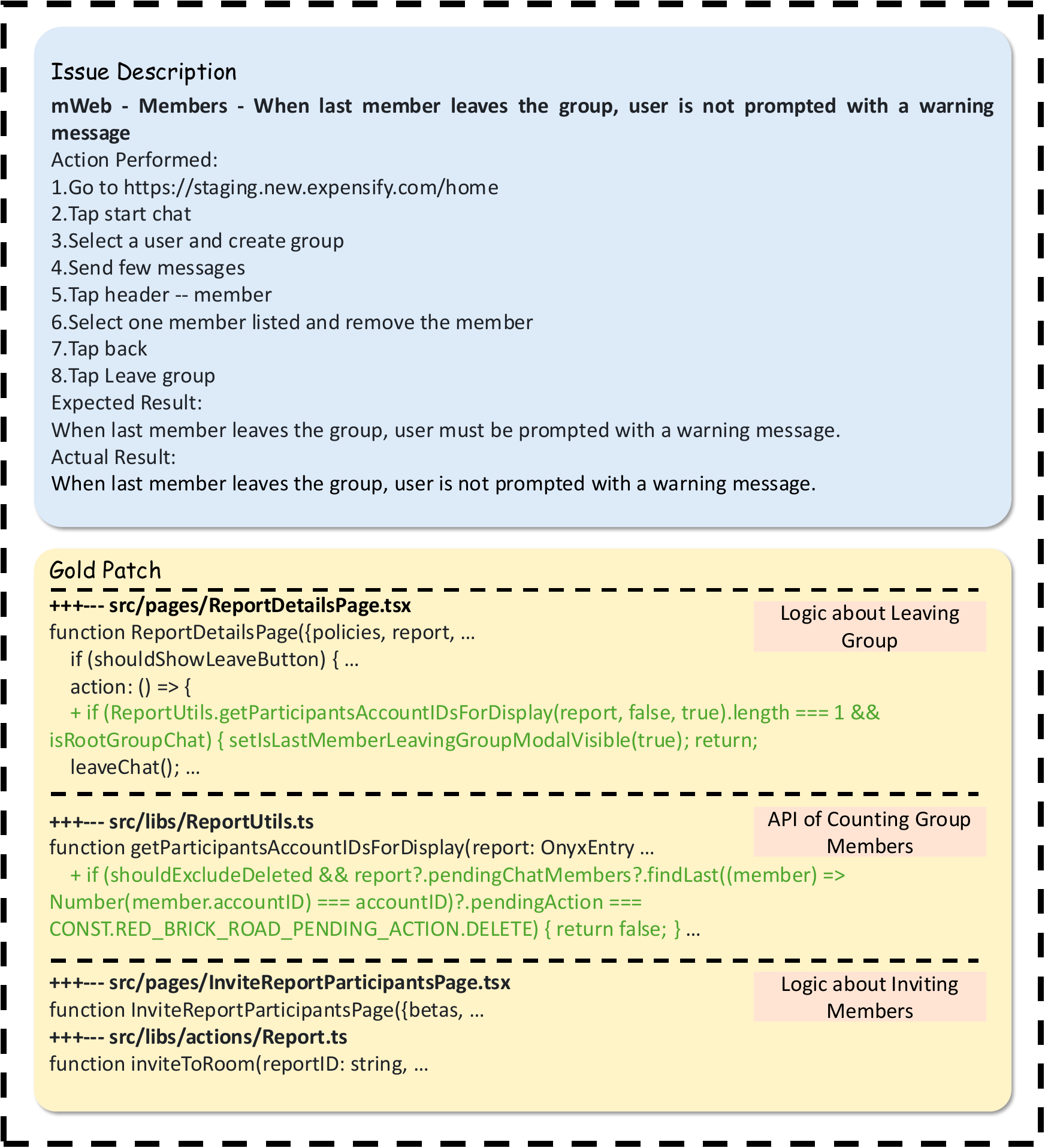}
        \caption{Issue Description and Gold Patch}\label{fig:motivation-issue}
    \end{subfigure}
    \hfill
    \begin{subfigure}[b]{0.49\textwidth}
        \includegraphics[width=1\linewidth]{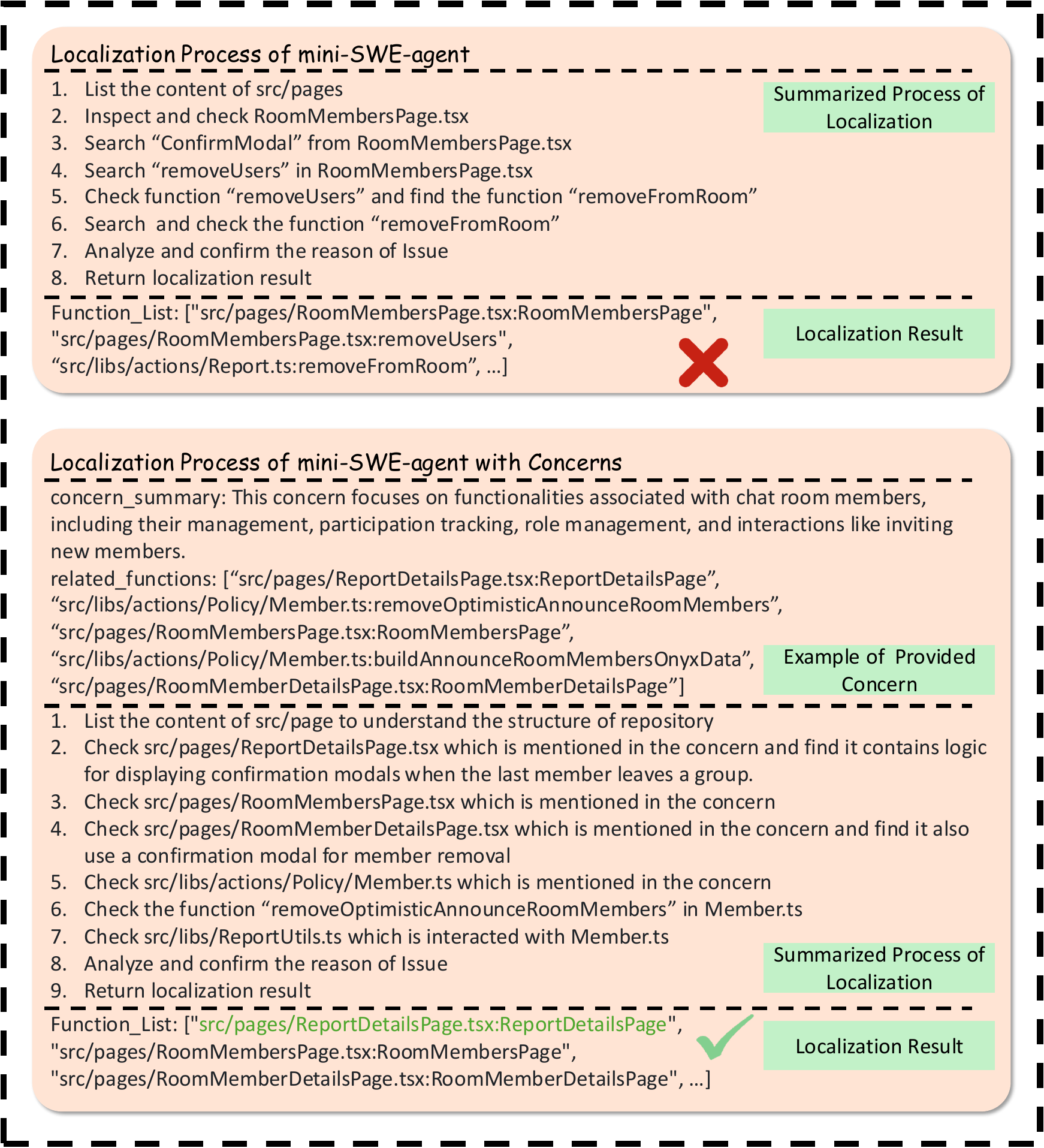}
        \caption{Comparison of Different Localization Processes}\label{fig:motivation-comparison}
    \end{subfigure}
    \caption{Motivating Example}
    \label{fig:motivation}
\end{figure}

In large-scale code repositories, issue localization is often hindered by the absence of a \textit{high-level conceptual view}, as well as by \textit{concern tangling} and \textit{concern scattering}, which make it difficult for existing methods to link abstract issue descriptions to the specific code elements responsible for the problem.
Figure \ref{fig:motivation} illustrates these challenges using a motivating example from a real issue in the Expensify repository. Figure \ref{fig:motivation-issue} shows the issue along with its corresponding fix patch. Figure \ref{fig:motivation-comparison} presents a comparison of the analysis processes of mini-SWE-agent~\cite{DBLP:journals/corr/abs-2405-15793} under different settings.

\subsection{Challenges in Issue Localization}

\textbf{Lack of High-level Conceptual View.} The entry point for analyzing the fix to this issue lies in the \inlinecode{ReportDetailsPage} function within \inlinecode{src/pages/ReportDetailsPage.tsx}, which is identified as the top file in the patch shown in Figure~\ref{fig:motivation-issue}. This function contains the direct logic—the call to \inlinecode{leaveChat()}—that triggers the confirmation modal when the last member leaves a group. At first glance, this might make the issue seem straightforward to localize. However, in practice it is not. For example, searching the repository with the keywords \textit{leave}, \textit{member}, and \textit{group} returns 19, 137, and 139 matched functions, respectively. The true concerns related to the issue are buried among a massive number of low-level code elements, with no high-level conceptual guidance to direct either human developers or LLM-based localization methods through the codebase.

\textbf{Concern Tangling Challenge.} Within \inlinecode{ReportDetailsPage}, the code directly related to the issue, the call to \inlinecode{leaveChat()}, occupies only a tiny fraction of the function, which spans 792 lines and encompasses many other report-related concerns, such as rendering the report header and managing report-specific actions. As a result, the leave-group logic is entangled with numerous unrelated responsibilities. Without explicit separation of concerns, isolating this small but critical piece of functionality becomes difficult, exemplifying the challenge of concern tangling.

\textbf{Concern Scattering Challenge.} After identifying the entry-point function \inlinecode{ReportDetailsPage}, fixing this issue required modifications across multiple files, each involving different functions that collectively implement the leave-group behavior. For example, handling the leave action and displaying the confirmation modal requires invoking the \inlinecode{getParticipantsAccountIDsForDisplay} function defined in \textit{ReportUtils.ts}. This utility computes the number of group members and determines whether the departing user is the last member, making it central to the fix. This highlights that effective issue analysis often requires aggregating scattered functionalities into a higher-level view. Beyond these two core functions, the patch also modifies \inlinecode{inviteToRoom} and \inlinecode{InviteReportParticipantsPage}, which deal with member invitations (Figure~\ref{fig:motivation-issue}). The former updates membership state by cleaning up pending members and refreshing participant lists, while the latter represents a separate invitation scenario that also relies on \inlinecode{getParticipantsAccountIDsForDisplay} to maintain consistency when removing users. These scattered functions jointly determine the correct behavior of leaving a group. This distribution of semantically related logic across multiple functions exemplifies concern scattering and underscores the difficulty of issue localization.

\subsection{Approach Comparison and Our Technique}
\textbf{Limitations of Existing Approaches.} Due to the challenges, existing automatic localization approaches often suffer from inefficient and ineffective exploration of the codebase. Figure~\ref{fig:motivation-comparison} illustrates the localization processes of mini-SWE-agent under two different settings, with the top one representing the original workflow. As shown, the agent identifies \inlinecode{RoomMembersPage} and \inlinecode{removeUsers} in \textit{RoomMembersPage.tsx} as the targets, following a sequence of file listings and code searches and inspections. However, lacking high-level guidance, the agent pursues an incorrect direction and becomes overly focused on low-level code details, ultimately leading to mislocalization. 

\textbf{Our Concern-Enhanced Idea.} To address these challenges, we propose abstracting conceptual knowledge by first decomposing fine-grained functionalities from original code functions (mitigating concern tangling), and then recomposing them into high-level conceptual concerns (mitigating concern scattering). During issue localization, the most relevant concerns are selected to provide high-level guidance, enabling more efficient and effective analysis of the issue and exploration of the codebase (providing high-level conceptual overview). As illustrated in the bottom process of Figure~\ref{fig:motivation-comparison}, once the constructed concerns are provided, mini-SWE-agent can correctly identify the entry-point function \inlinecode{ReportDetailsPage} by leveraging the associated concern.

\section{APPROACH}\label{sec-method}
We propose a novel approach, \tool, to enhance issue localization by abstracting and leveraging conceptual knowledge from code repositories. Rather than treating the repository as a flat collection of files/functions, \tool constructs a higher-level representation in the form of concerns—semantically coherent functionality clusters that serve as explicit clues during localization.

\begin{figure}
    \centering
    \includegraphics[width=\columnwidth]{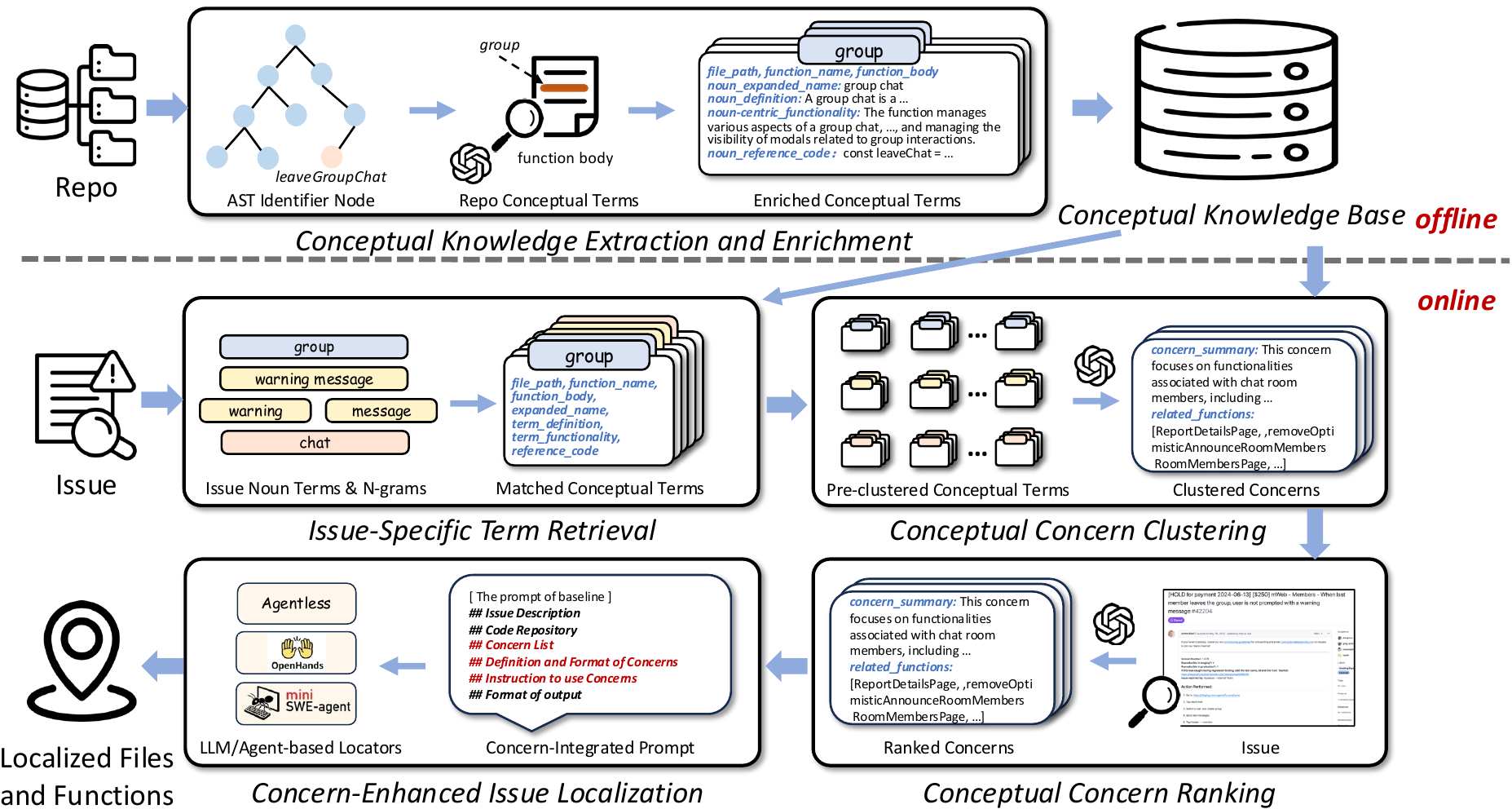}
    \caption{Overview of \tool}
    \label{fig:overview}
\end{figure}

As illustrated in Figure~\ref{fig:overview}, \tool operates in two stages. The offline stage performs Conceptual Knowledge Extraction and Enrichment, decoupling fine-grained functionalities associated with conceptual terms from individual functions and constructing a conceptual knowledge base for a given repository. The online stage consists of four steps: (i) Issue-Specific Term Retrieval, which matches conceptual terms from the knowledge base using keywords extracted from the issue; (ii) Conceptual Concern Clustering, which groups retrieved terms and their functionalities into semantically coherent concerns; (iii) Conceptual Concern Ranking, which filters and sorts concerns by relevance to the issue; and (iv) Concern-Enhanced Issue Localization, which incorporates the ranked concerns into issue localization tools through minimally intrusive prompt modifications.

\subsection{Conceptual Knowledge Extraction and Enrichment}
Given a code repository, \tool first extracts all conceptual terms from identifiers and explains them using formal names, definitions, and functionalities derived from the code context.

\subsubsection{Term Extraction}
Noun terms form the foundation for understanding a codebase because they represent domain-specific concepts or functional entities in source code, serving as a semantic bridge between natural language issue descriptions and code elements. A conceptual term is defined as a noun word or noun phrase that may include adjective modifiers or qualifiers (e.g., in \textit{personal bank account}, \textit{personal} is the adjective modifier and \textit{bank account} is the noun phrase).

Since conceptual terms are primarily embedded in identifiers such as function names, method names, and variable names, \tool extracts them by first parsing each source file into an Abstract Syntax Tree (AST) using Tree-sitter~\cite{treesitter}. From the AST, \tool identifies function definitions by locating nodes of type \texttt{function}. Within each function node, it collects all identifiers from \texttt{identifier} nodes. Each identifier is then split into a sequence of words according to camelCase and snake\_case naming conventions. The words are annotated with part-of-speech (POS) tags and grouped into chunks using POSSE~\cite{gupta2013part}. Based on POS tags and chunking, \tool extracts conceptual terms from identifiers. For example, the identifier \texttt{getUserNameById} is split, POS-tagged, and chunked into [$V$:\textit{get}, $N$:\textit{user name}, $P$:\textit{by}, $N$:\textit{id}], where $V$, $N$, and $P$ represent verb, noun, and preposition, respectively. From this, \textit{user name} and \textit{id} are extracted as noun terms. \tool then applies post-processing to lemmatize the terms using spaCy~\cite{spacy}, such as converting plural nouns to their singular form.
Finally, terms extracted from the same function are deduplicated, since repeated mentions of a term within a function context typically convey the same meaning.

\subsubsection{Term Explanation}
Since the extracted terms contain only literal information, \tool enriches their semantics by explaining them from multiple dimensions, leveraging the powerful code understanding capabilities of large language models (LLMs)~\cite{gpt4o, gpt4.1}. 

\textbf{Expanded Name.} A concept often appears in a codebase under different forms (aliases), such as abbreviations or shortened names. These variations can hinder both the construction of associations between issues and code elements and the tracing of connections among code elements, even though they refer to the same underlying concept. To address this, \tool attempts to expand each term into its formal name when possible, based on the surrounding function context. For example, \textit{req} and \textit{id} may be expanded to \textit{request} and \textit{transaction id}, respectively, depending on the function in which they appear. The instructions used to prompt the LLM for name expansion are as follows.
    \begin{mdframed}[linecolor=myblue!50,linewidth=2pt,roundcorner=10pt,backgroundcolor=gray!20,leftmargin=0pt,rightmargin=0pt,innertopmargin=2pt,innerbottommargin=2pt,innerleftmargin=2pt,innerrightmargin=2pt]
    \small
    \textsc{\textbf{Term Name Expansion Instructions:}}
    \begin{itemize}[label=-,leftmargin=10pt]
        \item Determine whether the given noun term is an abbreviation. 
        \item If it is, expand it into its full form based on the code context.
        \item After obtaining the full form, add necessary modifiers or qualifiers based on the context to make the noun term more specific and precise. 
        \item The final result must remain a concrete noun term (not a descriptive sentence), and use spaces between words within the term. 
    \end{itemize}
    \end{mdframed}

\textbf{Term Definition.} To further clarify the meaning of a term, a definition is required to describe the concept it represents. Importantly, the definition should focus on \textit{what} the term is, as inferred from the function context, rather than \textit{how} it is used in the function code. The instructions provided to the LLM for generating definitions are as follows.
    \begin{mdframed}[linecolor=myblue!50,linewidth=2pt,roundcorner=10pt,backgroundcolor=gray!20,leftmargin=0pt,rightmargin=0pt,innertopmargin=2pt,innerbottommargin=2pt,innerleftmargin=2pt,innerrightmargin=2pt]
    \small
    \textsc{\textbf{Term Definition Generation Instructions:}}
    \begin{itemize}[label=-,leftmargin=10pt]
        \item Provide a concise natural language sentence that defines what the given term **is**.  
        \item The definition should capture the intrinsic meaning of the noun term, independent of how it is used in the code. 
    \end{itemize}
    \end{mdframed}

\textbf{Term-Centric Functionality.} Code functionality is essential for analyzing the relationships between code elements and issues. However, as discussed in Section~\ref{sec-motiv}, coarse-grained, function-level summaries cannot adequately capture the multiple concerns that are often intertwined within complex functions. To address this, we propose a fine-grained, term-centric functionality summarization approach, which focuses exclusively on the logic directly related to each term rather than the entire function context. This enables \tool to decouple the concerns within a function based on the terms it involves, thereby providing more precise clues for associating issues with code elements. The instructions used to prompt the LLM are as follows.
    \begin{mdframed}[linecolor=myblue!50,linewidth=2pt,roundcorner=10pt,backgroundcolor=gray!20,leftmargin=0pt,rightmargin=0pt,innertopmargin=2pt,innerbottommargin=2pt,innerleftmargin=2pt,innerrightmargin=2pt]
    \small
    \textsc{\textbf{Term-Centric Functionality Summarization Instructions:}}
    \begin{itemize}[label=-,leftmargin=10pt]
        \item Analyze the functional semantics/responsiblities or business logic in the given function that specifically relates to the noun term. 
        \item Focus only on the part of the logic that is relevant to the noun term, **not the overall functionality of the entire function**. 
    \end{itemize}
    \end{mdframed}

\textbf{Reference Code.} \tool leverages the LLM to extract code snippets directly relevant to understanding a term from the entire function context, using the instructions shown below. Note that supplemental examples, such as style spreading or simple JSX wrappers, are included because the current experiments primarily involve web projects and JavaScript/TypeScript code. It is straightforward to add additional illustrations for other programming languages.
    \begin{mdframed}[linecolor=myblue!50,linewidth=2pt,roundcorner=10pt,backgroundcolor=gray!20,leftmargin=0pt,rightmargin=0pt,innertopmargin=2pt,innerbottommargin=2pt,innerleftmargin=2pt,innerrightmargin=2pt]
    \small
    \textsc{\textbf{Reference Code Extraction Instructions:}}
    \begin{itemize}[label=-,leftmargin=10pt]
        \item Identify and extract 1-3 concise code snippets (each 1-5 lines) that illustrate the term-centric functionality.
        \item Ensure the extracted snippets **directly** support understanding of the term-centric functionality, and exclude trivial or non-essential code (e.g., style spreading or simple JSX wrappers).
    \end{itemize}
    \end{mdframed}

\begin{figure}[ht!]
    \centering
    \includegraphics[width=\columnwidth]{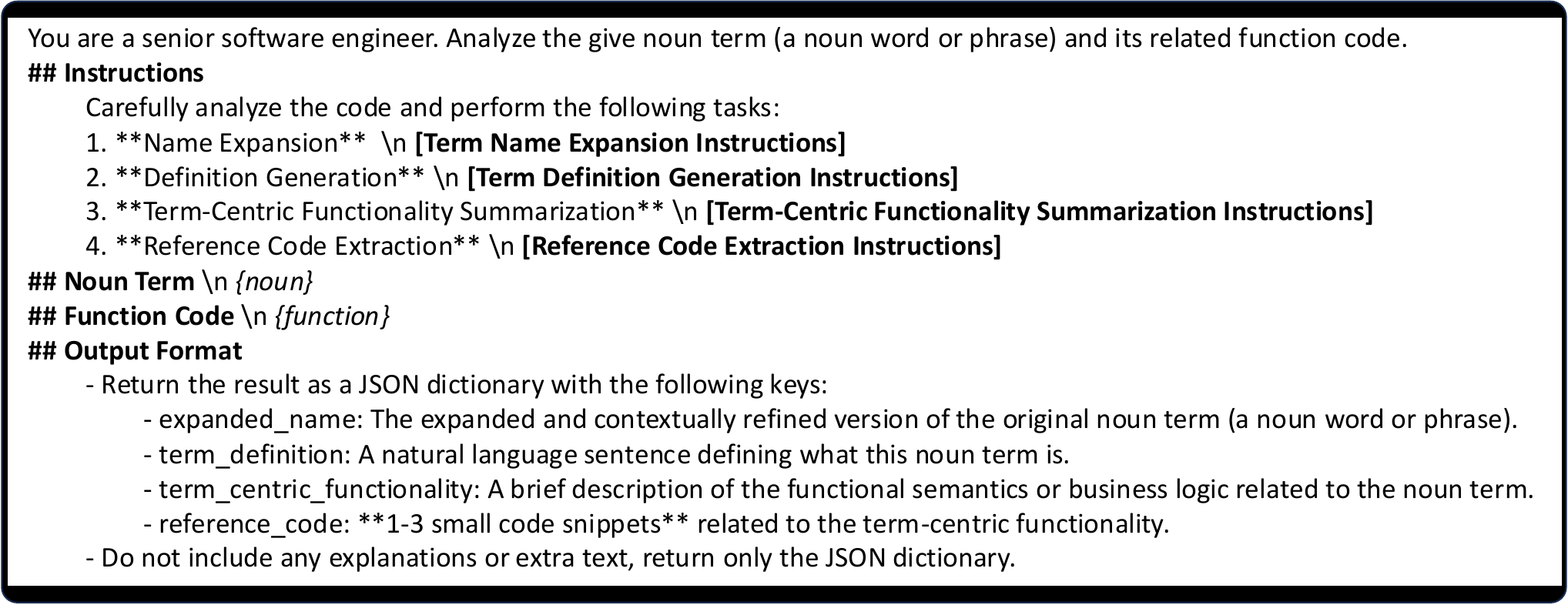}
    \caption{Prompt Template for Term Explanation}
    \label{fig:prompt-term-explanation}
\end{figure}

Because the above four types of explanatry information are relevant and dependent, we integrate them into one prompt with step-by-step chain-of-thoughts (CoT)~\cite{DBLP:conf/nips/Wei0SBIXCLZ22} instructions. This can ensure the coherence among the generated explanations, while reducing the complexity and cost of requesting the LLM. The integrated prompt template is as follows. The integrated prompt template is shown in Figure \ref{fig:prompt-term-explanation}.

\vspace{-3mm}

\subsubsection{Knowledge Base Construction}
\begin{figure}[ht!]
    \centering
    \includegraphics[width=\columnwidth]{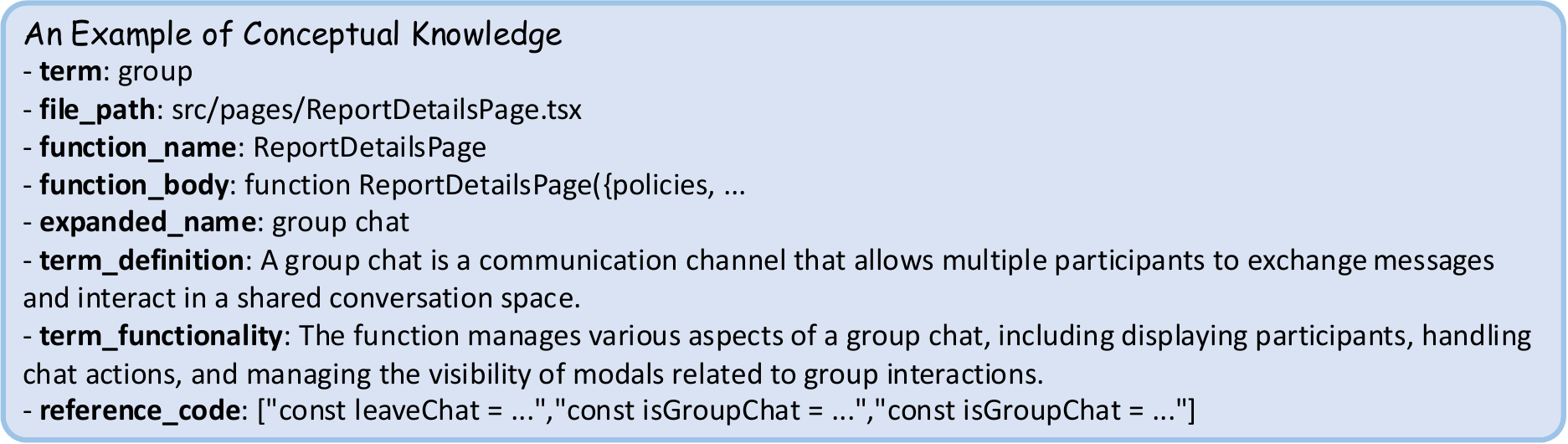}
    \caption{An Example of Extracted Conceptual Term and Its Explanations}
    \label{fig:example-enrich}
\end{figure}

The extracted conceptual terms, along with their explanations and metadata (such as the file paths and function names from which the terms were extracted), are organized into a knowledge base that provides fine-grained conceptual knowledge of the code repository. An example of conceptual knowledge is shown in Figure \ref{fig:example-enrich}.

\subsection{Issue-Specific Term Retrieval}

Given an issue, \tool first extracts noun keywords from its title and then uses these keywords to retrieve related conceptual terms from the knowledge base.

\subsubsection{Keyword Extraction}
Since issue titles are written in natural language, \tool uses an LLM to extract noun words and phrases from them.
A post-processing step performs lemmatization and deduplication, producing a unique list of noun keywords that capture the key entities and concepts expressed in the issue. We restrict keyword extraction to issue titles because they are concise and focused on the main topic. In contrast, issue bodies are often less structured and may introduce irrelevant terms, adding noise to the retrieval process.

\subsubsection{Term Matching}
Using the extracted keywords, \tool retrieves relevant conceptual terms from the knowledge base. Specifically, the keywords are first decomposed into their n-gram variants. This n-gram–based strategy supports partial matching, enabling multi-word or partially overlapping phrases to be matched, thereby improving recall. For each n-gram, if it matches any conceptual term in the knowledge base, the matched terms are retained along with their metadata and explanations.
Terms matched by n-grams derived from the same keyword are grouped into a single term set. For example, if the keyword is \textit{warning message}, all terms matching \textit{warning}, \textit{message}, and \textit{warning message} are grouped into the same term set, as they are likely semantically related within the issue context.

\subsection{Conceptual Concern Clustering}
As discussed in Section~\ref{sec-motiv}, accurately analyzing the association between an issue and fine-grained functionalities in the codebase requires gathering scattered concerns related to the issue into a unified view. To this end, \tool performs concern clustering on the retrieved terms with their functionalities to produce semantically coherent concern clusters. 
The clustering process consists of two stages: similarity-based term pre-clustering followed by LLM-based concern clustering.

\subsubsection{Similarity-based Term Pre-Clustering}
A straightforward approach is to directly employ the LLM for clustering by leveraging its strong language processing and semantic understanding capabilities. However, this approach faces two key limitations. First, it fails to account for structural relationships behind the terms in the code, such as function call dependencies. Second, when the candidate set to be clustered is large (e.g., more than 100 functionalities), the LLM often omits a subset of candidates, leading to incomplete clusters.

To address these issues, for each term set retrieved by the same keyword from an issue, \tool employs a hybrid pre-clustering strategy. This method partitions the candidate terms into smaller subsets, each constrained by a predefined size limit (e.g., 100). Specifically, if the size of a term set exceeds the limit, pre-clustering is carried out using a standard hierarchical clustering algorithm~\cite{DBLP:journals/jacm/Cohen-AddadKMM19}. To execute the algorithm, the similarity between each pair of terms in the set, denoted as $nt_i$ and $nt_j$, is computed as follows: 
\textit{Sim($nt_{i}$,$nt_{j}$)} = $\frac{(name\_sim + def\_sim + func\_sim + call\_bonus)}{4}$,
where $name\_sim$, $def\_sim$, and $func\_sim$ represent the embedding similarities of terms with respect to their expanded names, definitions, and functionality, respectively. The embeddings for these explanations are generated using the sentence-transformer
model~\cite{DBLP:conf/emnlp/ReimersG19}, and the similarities are computed with cosine similarity. The $call\_bonus$ is set to 1 if the functions containing the two terms have a call relationship, and 0 otherwise.

After computing pairwise similarities among all terms in the set, we apply hierarchical clustering to group them. The similarity between two groups is defined as the average similarity of the terms they contain. At each iteration, the two most similar groups are merged, and this process continues until a hierarchical tree is built. We then traverse the tree in a top-down manner, selecting subtrees whose total number of terms does not exceed the predefined limit, and take these as the final clusters. The resulting subsets of terms are thus likely to share similar meanings or related functionalities.


\subsubsection{LLM-based Concern Clustering}
For each pre-clustered term subset, \tool employs an LLM to group the terms' functionalities into concerns, thereby capturing the functional aggregation relationships among them. The LLM first interprets the \hyperlink{concern-def}{\textsc{Concern Definition}}, then examines the provided atomic functionalities (associated with the terms), and identifies their conceptual and semantic relationships to aggregate them into concerns that reflect specific responsibilities or areas of interest. Each concern is accompanied by a summary that captures its conceptual semantics.

\begin{figure}
    \centering
    \includegraphics[width=\columnwidth]{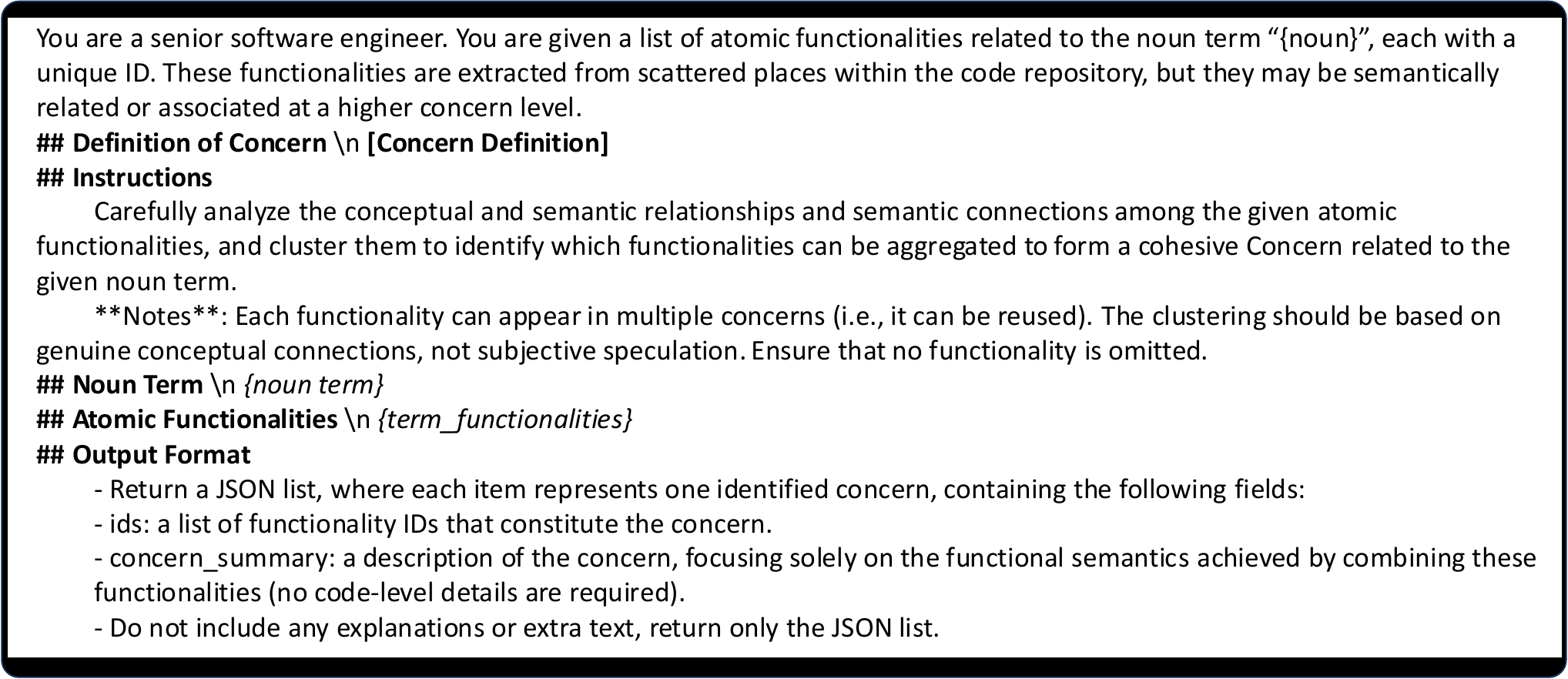}
    \caption{Prompt Template for Concern Clustering}
    \label{fig:prompt-concern-clustering}
\end{figure}

Importantly, an atomic functionality may belong to multiple concerns, as the design of reusable functionalities is a fundamental principle in software engineering. To mitigate the risk of the LLM hallucinating unreliable concerns, we incorporate instructions that constrain the model from engaging in subjective speculation. Building on these ideas and principles, the prompt to conduct the concern clustering operation is shown in Figure \ref{fig:prompt-concern-clustering}.



\begin{figure}
    \centering
    \includegraphics[width=\columnwidth]{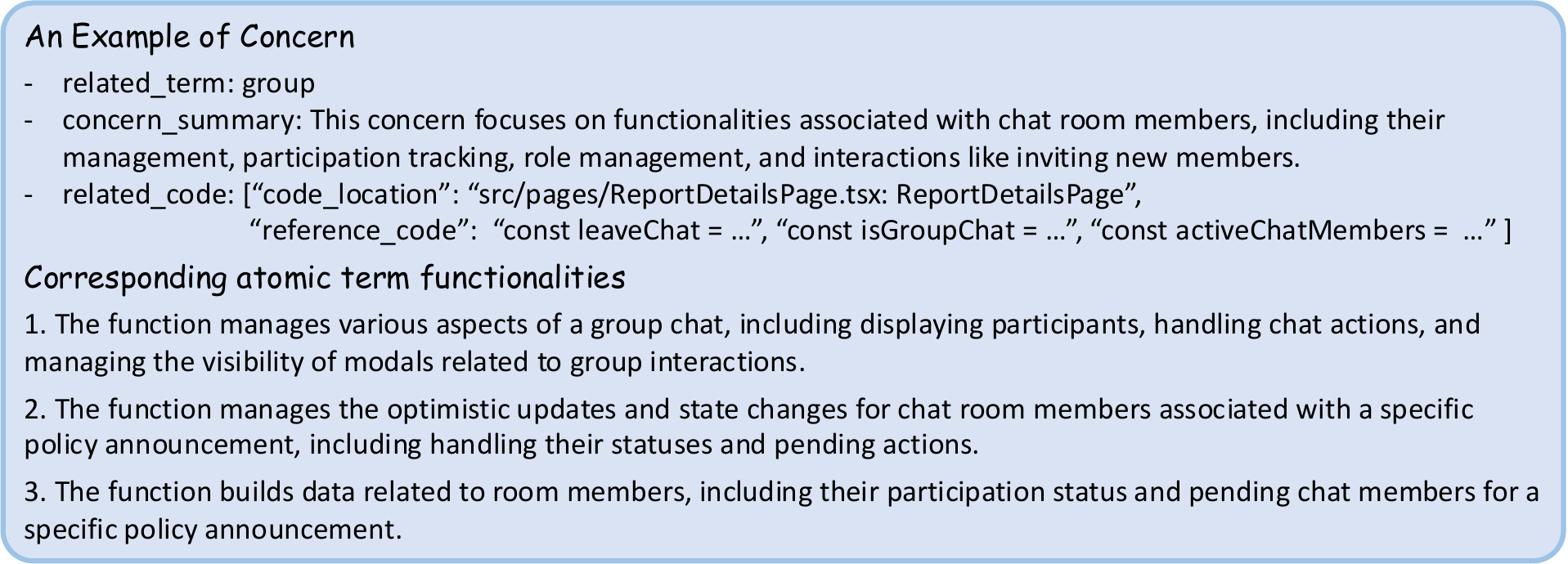}
    \caption{An Example of Concern and Its Corresponding Atomic Functionalities}
    \label{fig:example-concern}
\end{figure}

The clustered concerns are organized as follows. Note that atomic functionalities are excluded from the subsequent process because concerns provide higher-level semantic information that links multiple code locations. Including atomic functionalities may cause the model to focus on local details rather than adopting the broader perspective offered by concerns. An example of concern and its corresponding atomic term functionalities is shown in Figure \ref{fig:example-concern}.

\begin{mdframed}
[linecolor=myblue!50,linewidth=2pt,roundcorner=10pt,backgroundcolor=gray!20,leftmargin=0pt,rightmargin=0pt,innertopmargin=2pt,innerbottommargin=2pt,innerleftmargin=2pt,innerrightmargin=2pt]
\small
\hypertarget{concern-def}{\textsc{\textbf{Concern Format:}}} 
\begin{itemize}[label=-,leftmargin=15pt]
    \item related\_term: the noun term that these concerns are associated with.
    \item concern\_summary: a description of the concern, summarizing the functional semantics or business logic it represents.
    \item related\_code: a list of code snippets related to this concern, where each item contains:
        \begin{itemize}[label=-,leftmargin=15pt]
            \item code\_location: the file path and function name where the code is located \\(format: \texttt{file\_path:function\_name}).
            \item reference\_code: the actual code snippets relevant to this concern.
        \end{itemize}
\end{itemize}
\end{mdframed}

\subsection{Candidate Concern Ranking}
After concern clustering is performed on all retrieved term sets, multiple concerns may be produced for a given issue. Before feeding these concerns into the issue localization process, \tool ranks them based on their relevance to the issue. This ranking step is critical because issue localization methods, particularly LLM-agent approaches, struggle to maintain attention and extract precise clues when presented with an excessive number of concerns. The ranking process is therefore divided into two stages: (i) similarity-based concern selection and (ii) LLM-based concern reranking.

\subsubsection{Similarity-based Concern Selection}
In the first stage, \tool leverages embedding-based semantic similarity to filter the most relevant concerns. Specifically, it encodes both the issue title and each concern summary into vector representations using a pre-trained embedding model~\cite{DBLP:conf/emnlp/ReimersG19}. The cosine similarity between the issue title vector and each concern vector is then computed. Based on these scores, \tool selects the top-50 concerns with the highest similarities as candidates for further refinement. This step serves as a lightweight pre-filtering mechanism, ensuring efficiency by narrowing down the candidate space while preserving semantic relevance.

\subsubsection{LLM-based Concern Reranking}
\begin{figure}
    \centering
    \includegraphics[width=\columnwidth]{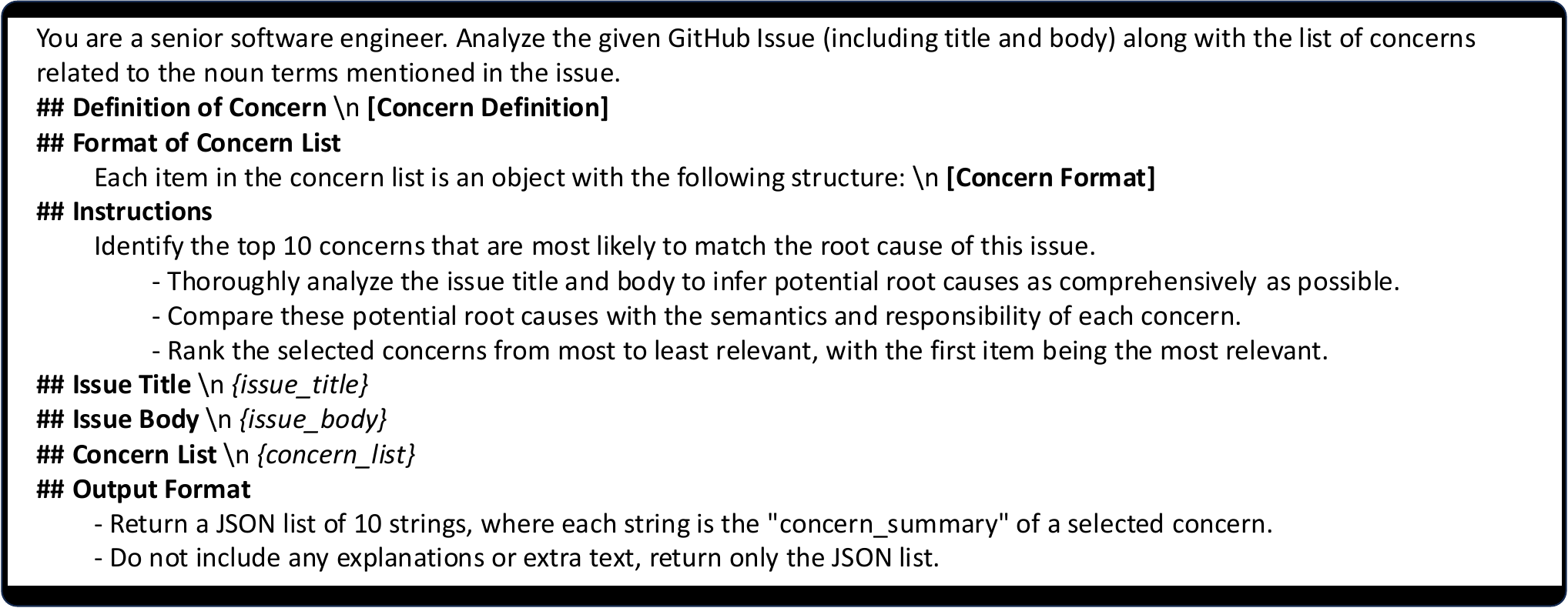}
    \caption{Prompt Template for Concern Ranking}
    \label{fig:prompt-concern-ranking}
\end{figure}

In the second stage, \tool performs a finer-grained ranking using an LLM. The issue title, issue body, and detailed information of the top-50 candidate concerns are provided as input. The LLM evaluates the relevance of each concern to the issue, identifies those most likely to represent the root cause, and produces a reranked list. Finally, the top-$N$ concerns (10 in the current implementation) are retained as the final ranking output. The prompt used to rank concerns is shown in Figure~\ref{fig:prompt-concern-ranking}.

\subsection{Concern-Enhanced Issue Localization}
The ranked concerns from \tool are leveraged to enhance LLM-powered issue localization tools, whether workflow-based (e.g., AgentLess~\cite{DBLP:journals/corr/abs-2407-01489}) or agent-based (e.g., OpenHands~\cite{DBLP:conf/iclr/0001LSXTZPSLSTL25}). This enhancement is achieved by integrating concern-related contexts and instructions into the tools' prompts in a minimally intrusive manner, maximizing the preservation of their original behavior.

\subsubsection{Basic Principles}
The key principle is to guide the tools to infer potential root causes of a given issue and establish associations with the provided concerns. By referencing the concerns, the tools can narrow their search and analysis scope from the entire repository to concern-relevant code. At the same time, to mitigate the risk of misleading or noisy concerns, we preserve the autonomous exploration capabilities of the agent-based tools (e.g., searching code elements with \texttt{grep}) rather than restricting them solely to concern-based exploration.

\subsubsection{Concrete Operations}
Specifically, three components about concern—\textsc{Concern Definition}, \textsc{Concern Format}, and \textsc{Concern List}—are added to all localization tools' prompts. In addition, we insert customized concern-related instructions depending on the type of tool:
\begin{itemize}[leftmargin=15pt]
    \item \textit{Workflow-based Locators.} Approaches such as AgentLess, which typically perform hierarchical traversal from folders to files to functions, are adapted by replacing the original instruction so that LLMs select candidate files or functions by traversing the concerns. For example, in the prompt of AgentLess, the original instruction \textit{``You can only search for files within the current given file repository structure''} is extended with \textit{``Pay special attention to files that appear in the Concern list and are semantically connected to the problem description''}.
    
    \item \textit{Agent-based Locators.} These tools, such as OpenHands and mini-SWE-agent, are typically more autonomous in exploring and analyzing the codebase. For them, we add additional concern-reference instructions to their system prompts, encouraging LLMs to use concerns as contextual clues during codebase exploration. For example, in the prompt of OpenHands, the original instruction \textit{``Locate specific files and functions requiring changes or containing critical information for resolving the issue''} is extended with \textit{``using the provided concerns (descriptions, files, and functions) as guiding context''}.

\end{itemize}

\section{Evaluation}\label{sec-data}

To evaluate the effectiveness and robustness of the proposed Concern-enhanced issue localization framework, we formulate the following research questions:

\begin{itemize}[leftmargin=*]
    \item \textbf{Effectiveness (RQ1):} How effective is \tool in improving issue localization compared to existing methods?
    \item \textbf{Generalizability (RQ2):} Can \tool generalize across different base LLMs?
    \item \textbf{Ablation Study (RQ3):} How does the performance change when replacing concern-based representations in \tool with simpler alternatives?
    \item \textbf{Reliability of Concerns (RQ4):} To what extent do the generated concerns align with human judgment while demonstrating consistency and reliability?
\end{itemize}

\subsection{Setup}

\subsubsection{Benchmark}


We adopt SWE-Lancer Diamond~\cite{DBLP:journals/corr/abs-2502-12115} as the foundation of our benchmark. SWE-Lancer Diamond is an open-source evaluation subset derived from SWE-Lancer~\cite{DBLP:journals/corr/abs-2502-12115}, which comprises over 1,400 freelance software engineering tasks (issues) with total real-world payouts exceeding \$1 million USD. The benchmark is well-suited for the issue localization task. Specifically, SWE-Lancer Diamond includes real-world software engineering tasks drawn from Expensify~\cite{expensify}, an expense management system with a large-scale codebase of 2.04 million code lines. The codebase encompasses rich conceptual knowledge (e.g., expense) and multiple interdependent functionalities (e.g., management of money request). This scale and complexity make it particularly suitable for evaluating issue localization under conceptual challenges. In contrast, we did not adopt SWE-Bench~\cite{DBLP:conf/iclr/JimenezYWYPPN24}, as it primarily consists of isolated Python Library projects. SWE-Lancer Diamond includes 237 independent engineering issues, each accompanied by a bug reintroduction procedure linked to a specific repository commit. The benchmark evaluates issue-fixing methods by reverting the repository to the specific commit and executing the bug reintroduction procedure to restore the repository to the state of the issue.


We process SWE-Lancer Diamond into a dataset tailored for file-level and function-level issue localization, denoted as \textbf{SWE-Lancer-Loc}. Specifically, we retain 216 issues and filter out 21 issues whose bug reintroduction procedures are faulty or unreproducible. To establish the file-level and function-level localization ground truth, we compare the repository states before and after executing the bug reintroduction procedure. The relevant changed files and functions are recorded as the issue’s localization targets, represented as \textit{gold\_files} and \textit{gold\_functions}. Therefore, each issue includes the following fields for both file-level and function-level issue localization:

\begin{itemize}[leftmargin=*]
    \item \textit{issue\_description}: The textual description of the issue, including both the title and body.  
    \item \textit{repository}: The code repository in which the issue occurs.  
    \item \textit{commit\_id}: The specific commit of the repository used to inject the issue.  
    \item \textit{bug\_reintroduce.patch}: The script that injects the issue into the repository at the specified commit.  
    \item \textit{gold\_files}: A list of file paths representing the ground truth for file-level issue localization.  
    \item \textit{gold\_functions}: A list of functions that serve as the ground truth for function-level issue localization, represented in the format \texttt{file\_path:function\_name}.
\end{itemize}

\subsubsection{Baselines}

We compare our approach against three representative baselines:

\begin{itemize}[leftmargin=*]
    \item \textbf{AgentLess~\cite{DBLP:journals/corr/abs-2407-01489}:} AgentLess hierarchically localizes issues by combining repository structure with embedding-based retrieval, and then incrementally narrows down from suspicious files to specific functions or lines using LLMs.

    \item \textbf{OpenHands~\cite{DBLP:conf/iclr/0001LSXTZPSLSTL25}:} OpenHands is an agent-based approach for repository-level issue analysis and localization, which extracts key problem information from issue, maps it to relevant modules,  and identifies specific code locations for modification.

    \item \textbf{mini-SWE-agent~\cite{DBLP:journals/corr/abs-2405-15793}:} mini-SWE-agent is a lightweight variant of SWE-agent that leverages LLM-driven reasoning and tool usage to localize issues, simulating developer actions such as reading repository structures and retrieving candidate files in a cost-efficient manner.
    
\end{itemize}

\subsubsection{Metrics}

To assess the effectiveness of issue localization at the file level (and respectively at the function level), we adopt \textbf{Hit@k} and \textbf{Recall@k} where $k \in \{1,5,10\}$.

\begin{itemize}[leftmargin=*]
    \item \textbf{Hit@k:} This metric measures whether at least one gold file (resp. function) is included within the top $k$ predictions. 
    \item \textbf{Recall@k:} This metric quantifies the proportion of gold files (resp. functions) that are correctly retrieved within the top $k$ predictions.
\end{itemize}



\subsubsection{Implementation}

All experiments are conducted with temperature fixed to 0 and a maximum of 30 interaction rounds per issue. While implementing \tool, we adopt a cost-aware strategy to balance effectiveness and efficiency when selecting LLMs for different stages of the framework. Specifically, we use the lightweight \textit{GPT-4o-mini} model for repository term enrichment, issue-specific term extraction, and concern ranking. For concern clustering, we employ the more capable \textit{GPT-4o} model to ensure higher reliability. When integrating \tool with baseline systems, we further evaluate three LLM configurations—\textit{GPT-4o}, \textit{GPT-4o-mini}, and \textit{GPT-4.1} under consistent inference settings. In RQ1 and RQ3, we adopt \textit{GPT-4o} for optimal performance, and in RQ2, we evaluate the performance using all three LLMs. 

To faithfully reproduce the state in which each issue occurs, we first restore the repository to the \textit{commit\_id} associated with the issue and then apply the corresponding \textit{bug\_reintroduce.patch} to inject the bug back into the codebase before performing localization.

\subsection{RQ1: Effectiveness}

To evaluate the effectiveness of \tool, we compare \tool against three baseline systems (i.e., AgentLess, OpenHands, and mini-SWE-agent) at both file-level and function-level. All baselines are implemented using \textit{GPT-4o} to ensure a fair comparison.

\begin{table*}[!t]
    \centering
    \footnotesize
    \renewcommand{\arraystretch}{1.1} 
    \setlength{\tabcolsep}{3pt}
    \caption{File-Level Localization Effectiveness}
    \begin{tabular}{lccccccc}   
    \toprule
    \multirow{2}{*}{\textbf{Method}} & \multicolumn{3}{c}{\textbf{Hit@k}} & & \multicolumn{3}{c}{\textbf{Recall@k}} \\ 
    \cmidrule{2-4} \cmidrule{6-8}
    & $k=1$ & $k=5$ & $k=10$ & & $k=1$ & $k=5$ & $k=10$ \\ 
    \midrule
    AgentLess & 15.28\%  & 33.80\% & 36.57\% & & 10.33\% & 22.98\% & 25.33\% \\ 
    \quad + \tool & 29.17\% (\textcolor{ForestGreen}{$90.90\%^\uparrow$})  
        & 51.85\% (\textcolor{ForestGreen}{$53.40\%^\uparrow$})  
        & 56.94\% (\textcolor{ForestGreen}{$55.70\%^\uparrow$})
        &
        & 19.27\% (\textcolor{ForestGreen}{$86.54\%^\uparrow$}) 
        & 37.01\% (\textcolor{ForestGreen}{$61.05\%^\uparrow$})  
        & 41.03\% (\textcolor{ForestGreen}{$61.98\%^\uparrow$})  \\
    \hline
    OpenHands & 18.98\% & 40.28\% & 44.44\% & & 13.06\% & 29.60\% & 33.76\% \\ 
    \quad + \tool & 19.91\% (\textcolor{ForestGreen}{$4.90\%^\uparrow$})  
        & 45.37\% (\textcolor{ForestGreen}{$12.64\%^\uparrow$})  
        & 55.56\% (\textcolor{ForestGreen}{$25.02\%^\uparrow$})  
        &
        & 13.52\% (\textcolor{ForestGreen}{$3.52\%^\uparrow$}) 
        & 32.53\% (\textcolor{ForestGreen}{$9.90\%^\uparrow$})  
        & 40.89\% (\textcolor{ForestGreen}{$21.12\%^\uparrow$})  \\
    \hline
    mini-SWE-agent & 24.07\% & 50.93\% & 56.02\% & & 17.34\% & 37.13\% & 44.21\% \\ 
    \quad + \tool & 25.46\% (\textcolor{ForestGreen}{$5.77\%^\uparrow$})  
        & 52.31\% (\textcolor{ForestGreen}{$2.71\%^\uparrow$})  
        & 64.35\% (\textcolor{ForestGreen}{$14.87\%^\uparrow$})  
        &
        & 18.86\% (\textcolor{ForestGreen}{$8.77\%^\uparrow$}) 
        & 37.37\% (\textcolor{ForestGreen}{$0.65\%^\uparrow$})  
        & 46.93\% (\textcolor{ForestGreen}{$6.15\%^\uparrow$})  \\
    \hline
    Average Improvement 
        & \textcolor{ForestGreen}{33.86\%$\bm{^\uparrow}$}  
        & \textcolor{ForestGreen}{22.92\%$\bm{^\uparrow}$} 
        & \textcolor{ForestGreen}{31.86\%$\bm{^\uparrow}$} 
        & 
        & \textcolor{ForestGreen}{32.94\%$\bm{^\uparrow}$}
        & \textcolor{ForestGreen}{23.87\%$\bm{^\uparrow}$} 
        & \textcolor{ForestGreen}{29.75\%$\bm{^\uparrow}$} \\ 
    \bottomrule
    \end{tabular}
    \label{tab:rq1:file}
\end{table*}

\textbf{Results.} Table \ref{tab:rq1:file} presents the effectiveness of \tool in file-level issue localization across the three baselines. Overall, \tool consistently enhances file localization effectiveness across all baselines and multiple k values regarding Hit@k and Recall@k. Across the three baselines, \tool achieves average relative Hit increases of 33.86\%, 22.92\%, and 31.86\% at k = 1, 5, and 10, respectively. Similarly, \tool also achieves average relative Recall scores of 32.94\%, 23.87\%, and 29.75\% at k = 1, 5, and 10, respectively. On OpenHands, \tool achieves the largest relative gains with a Hit increase of 25.02\% and a Recall increase of 21.12\% at k = 10. Table \ref{tab:rq1:function} presents the effectiveness of \tool in function-level issue localization across the three baselines. Overall, incorporating \tool significantly improves localization performance across all three baselines. On average, \tool improves Hit by 59.19\%, 51.21\%, and 60.17\% and Recall by 55.09\%, 47.42\%, and 56.88\% at $k = 1, 5,$ and $10$, respectively. For individual baselines, the largest improvements are observed in AgentLess, where Hit@1 increases by 127.18\% and Recall@1 increases by 108.90\%. OpenHands also benefits substantially, achieving a 65.83\% improvement at Hit@5 and a 66.21\% improvement at Recall@5.

\begin{table*}
    \centering
    \footnotesize
    \renewcommand{\arraystretch}{1.1}
    \setlength{\tabcolsep}{3pt}
    \caption{Function-Level Localization Effectiveness}
    \begin{tabular}{lccccccc}   
    \toprule
    \multirow{2}{*}{\textbf{Method}} & \multicolumn{3}{c}{\textbf{Hit@k}} & & \multicolumn{3}{c}{\textbf{Recall@k}} \\ 
    \cmidrule{2-4} \cmidrule{6-8}
    & $k=1$ & $k=5$ & $k=10$ & & $k=1$ & $k=5$ & $k=10$ \\ 
    \midrule
    AgentLess & 10.19\%  & 23.15\% & 25.00\% & & 8.31\%  & 19.14\% & 20.42\% \\ 
    \quad + \tool & 23.15\% (\textcolor{ForestGreen}{$127.18\%^\uparrow$})  
        & 37.50\% (\textcolor{ForestGreen}{$61.99\%^\uparrow$})  
        & 44.91\% (\textcolor{ForestGreen}{$79.64\%^\uparrow$})
        &
        & 17.36\% (\textcolor{ForestGreen}{$108.90\%^\uparrow$}) 
        & 29.27\% (\textcolor{ForestGreen}{$52.93\%^\uparrow$})  
        & 34.95\% (\textcolor{ForestGreen}{$71.16\%^\uparrow$})  \\
    \hline
    OpenHands & 9.72\% & 17.59\% & 21.76\% & & 7.56\% & 14.62\% & 17.72\% \\ 
    \quad + \tool & 14.35\% (\textcolor{ForestGreen}{$47.63\%^\uparrow$})  
        & 29.17\% (\textcolor{ForestGreen}{$65.83\%^\uparrow$})  
        & 35.19\% (\textcolor{ForestGreen}{$61.72\%^\uparrow$})  
        &
        & 11.80\% (\textcolor{ForestGreen}{$56.08\%^\uparrow$}) 
        & 24.30\% (\textcolor{ForestGreen}{$66.21\%^\uparrow$})  
        & 29.24\% (\textcolor{ForestGreen}{$65.01\%^\uparrow$})  \\
    \hline
    mini-SWE-agent & 16.67\% & 28.70\% & 31.94\% & & 13.79\% & 24.31\% & 27.01\% \\ 
    \quad + \tool & 17.13\% (\textcolor{ForestGreen}{$2.76\%^\uparrow$})  
        & 36.11\% (\textcolor{ForestGreen}{$25.82\%^\uparrow$})  
        & 44.44\% (\textcolor{ForestGreen}{$39.14\%^\uparrow$})  
        &
        & 13.83\% (\textcolor{ForestGreen}{$0.29\%^\uparrow$}) 
        & 29.93\% (\textcolor{ForestGreen}{$23.12\%^\uparrow$})  
        & 36.32\% (\textcolor{ForestGreen}{$34.47\%^\uparrow$})  \\
    \hline
    Average Improvement 
        & \textcolor{ForestGreen}{59.19\%$\bm{^\uparrow}$}  
        & \textcolor{ForestGreen}{51.21\%$\bm{^\uparrow}$} 
        & \textcolor{ForestGreen}{60.17\%$\bm{^\uparrow}$} 
        & 
        & \textcolor{ForestGreen}{55.09\%$\bm{^\uparrow}$}
        & \textcolor{ForestGreen}{47.42\%$\bm{^\uparrow}$} 
        & \textcolor{ForestGreen}{56.88\%$\bm{^\uparrow}$} \\ 
    \bottomrule
    \end{tabular}
    \label{tab:rq1:function}
\end{table*}

\textbf{Analysis.}
The results demonstrate that \tool consistently enhances localization performance across all three baselines. At the file level, \tool consistently improves localization effectiveness, yielding average relative Hit and Recall gains of more than 22\% across all $k$ values. At the function level, the improvements are even more pronounced, with average relative Hit and Recall gains exceeding 46\%. These substantial gains across different granularities and evaluation metrics confirm that \tool is able to effectively boost the localization capability of diverse baselines. Moreover, the improvements are more pronounced at the function level, since function-level localization is inherently more challenging, and our concerns explicitly include functions relevant to the issue, thereby offering fine-grained contextual signals that guide the baselines toward the correct function locations. The main reason behind these improvements lies in the quality of the concerns provided by \tool. Concerns act as high-quality contextual clues that are closely aligned with the issue description. To validate this, we calculate the average Hit score of gold files and functions achieved by the concerns, which are 0.62 and 0.50, respectively(see Table \ref{tab:rq3:compare}, for a detailed discussion). These values indicate that the extracted concerns consistently capture the ground-truth locations of issues, providing strong and relevant signals. With this additional guidance, baselines can directly focus on a narrowed search space rather than searching across the entire repository, thereby achieving significantly higher localization accuracy and efficiency.

\summary{\tool consistently improves localization performance across all three baselines, achieving average relative gains of over 22\% in Hit and 46\% in Recall across all $k$ values for file-level and function-level issue localization, respectively.}

\subsection{RQ2: Generalizability}\label{sec:rq2}

\begin{table*}[!t]
    \centering
    \footnotesize
    \renewcommand{\arraystretch}{1.1}
    \setlength{\tabcolsep}{2pt}
    \caption{File-Level Model Generalizability}
    \begin{tabular}{lccccccc}   
    \toprule
    \multirow{2}{*}{\textbf{Method}} & \multicolumn{3}{c}{\textbf{Hit@1}} & & \multicolumn{3}{c}{\textbf{Recall@10}} \\ 
    \cmidrule{2-4} \cmidrule{6-8}
    & GPT-4o & GPT-4o-mini & GPT-4.1 & & GPT-4o & GPT-4o-mini & GPT-4.1 \\ 
    \midrule
    AgentLess & 15.28\%  & 11.11\% & 25.93\% & & 25.33\% & 18.23\%  & 46.25\% \\ 
    \quad + \tool & 29.17\% (\textcolor{ForestGreen}{$90.90\%^\uparrow$})  
        & 19.44\% (\textcolor{ForestGreen}{$74.98\%^\uparrow$})  
        & 32.41\% (\textcolor{ForestGreen}{$24.99\%^\uparrow$})  
        &
        & 41.03\% (\textcolor{ForestGreen}{$61.98\%^\uparrow$}) 
        & 36.01\% (\textcolor{ForestGreen}{$97.53\%^\uparrow$})  
        & 52.55\% (\textcolor{ForestGreen}{$13.62\%^\uparrow$})  \\
    \hline
    OpenHands & 18.98\%  & 4.17\% & 1.39\% & & 33.76\% & 8.26\%  & 1.62\% \\ 
    \quad + \tool & 19.91\% (\textcolor{ForestGreen}{$4.90\%^\uparrow$})  
        & 9.72\% (\textcolor{ForestGreen}{$133.09\%^\uparrow$}) 
        & 3.70\% (\textcolor{ForestGreen}{$166.19\%^\uparrow$})  
        &
        & 40.89\% (\textcolor{ForestGreen}{$21.12\%^\uparrow$}) 
        & 20.69\% (\textcolor{ForestGreen}{$150.48\%^\uparrow$})  
        & 4.34\% (\textcolor{ForestGreen}{$167.90\%^\uparrow$})  \\
    \hline
    mini-SWE-agent & 24.07\%  & 8.33\% & 26.39\% & & 44.21\% & 14.93\%  & 40.15\% \\ 
    \quad + \tool & 25.46\% (\textcolor{ForestGreen}{$5.77\%^\uparrow$})  
        & 14.35\% (\textcolor{ForestGreen}{$72.27\%^\uparrow$}) 
        & 41.67\% (\textcolor{ForestGreen}{$57.90\%^\uparrow$})  
        &
        & 46.93\% (\textcolor{ForestGreen}{$6.15\%^\uparrow$}) 
        & 24.14\% (\textcolor{ForestGreen}{$61.69\%^\uparrow$})  
        & 55.74\% (\textcolor{ForestGreen}{$38.83\%^\uparrow$})  \\
    \bottomrule
    \end{tabular}
    \label{tab:rq2:file}
\end{table*}

\begin{table*}[!t]
    \centering
    \footnotesize
    \renewcommand{\arraystretch}{1.1}
    \setlength{\tabcolsep}{3pt}
    \caption{Function-Level Model Generalizability}
    \begin{tabular}{lccccccc}   
    \toprule
    \multirow{2}{*}{\textbf{Method}} & \multicolumn{3}{c}{\textbf{Hit@1}} & & \multicolumn{3}{c}{\textbf{Recall@10}} \\ 
    \cmidrule{2-4} \cmidrule{6-8}
    & GPT-4o & GPT-4o-mini & GPT-4.1 & & GPT-4o & GPT-4o-mini & GPT-4.1 \\ 
    \midrule
    AgentLess & 10.19\%  & 9.72\% & 15.74\% & & 20.42\% & 16.94\%  & 28.23\% \\ 
    \quad + \tool & 23.15\% (\textcolor{ForestGreen}{$127.18\%^\uparrow$})  
        & 15.74\% (\textcolor{ForestGreen}{$61.93\%^\uparrow$})  
        & 19.44\% (\textcolor{ForestGreen}{$23.51\%^\uparrow$})  
        &
        & 34.95\% (\textcolor{ForestGreen}{$71.16\%^\uparrow$}) 
        & 28.33\% (\textcolor{ForestGreen}{$67.24\%^\uparrow$})  
        & 29.03\% (\textcolor{ForestGreen}{$2.83\%^\uparrow$})  \\
    \hline
    OpenHands & 9.72\%  & 2.31\% & 0.46\% & & 17.72\% & 2.43\%  & 1.62\% \\ 
    \quad + \tool & 14.35\% (\textcolor{ForestGreen}{$47.63\%^\uparrow$})  
        & 5.56\% (\textcolor{ForestGreen}{$140.69\%^\uparrow$}) 
        & 2.78\% (\textcolor{ForestGreen}{$504.35\%^\uparrow$})  
        &
        & 29.24\% (\textcolor{ForestGreen}{$65.01\%^\uparrow$}) 
        & 11.57\% (\textcolor{ForestGreen}{$376.13\%^\uparrow$})  
        & 4.09\% (\textcolor{ForestGreen}{$152.47\%^\uparrow$})  \\
    \hline
    mini-SWE-agent & 16.67\%  & 6.02\% & 19.91\% & & 27.01\% & 7.75\%  & 29.81\% \\ 
    \quad + \tool & 17.13\% (\textcolor{ForestGreen}{$2.76\%^\uparrow$})  
        & 7.41\% (\textcolor{ForestGreen}{$23.09\%^\uparrow$}) 
        & 25.93\% (\textcolor{ForestGreen}{$30.24\%^\uparrow$})  
        &
        & 36.32\% (\textcolor{ForestGreen}{$34.47\%^\uparrow$}) 
        & 15.24\% (\textcolor{ForestGreen}{$96.65\%^\uparrow$})  
        & 42.70\% (\textcolor{ForestGreen}{$43.24\%^\uparrow$})  \\
    \bottomrule
    \end{tabular}
    \label{tab:rq2:function}
\end{table*}

To assess the generalizability of \tool, we evaluate its performance on three baselines using three LLMs, namely \textit{GPT-4o}, \textit{GPT-4o-mini}, and \textit{GPT-4.1}. We use \textbf{Hit@1} and \textbf{Recall@10} as evaluation metrics, since the top-ranked prediction is most critical in practice and Recall@10 better reflects coverage across candidate results.

\textbf{Results.} Table \ref{tab:rq2:file} reports the generalizability of \tool for file-level issue localization across different base models. The results show that \tool consistently brings improvements regardless of the different models, with relative Hit@1 gains ranging from 4.90\% to 166.19\% and Recall@10 gains between 6.15\% and 167.90\%. For AgentLess, both Hit@1 and Recall@10 achieve the highest values when using GPT-4.1 (32.41\% and 52.55\%, respectively). OpenHands performs best with GPT-4o, reaching 19.91\% Hit@1 and 40.89\% Recall@10, while performance drops on GPT-4o-mini and GPT-4.1. Table \ref{tab:rq2:function} further evaluates generalizability at the function level. Overall, \tool consistently improves function localization effectiveness regardless of the underlying model across all baselines. Specifically, for GPT-4o, GPT-4o-mini, and GPT-4.1, \tool yields average gains of 59.19\%, 75.24\%, and 186.03\% in Hit@1, and 56.88\%, 180.01\% and 66.18\% in Recall@10, respectively. For mini-SWE-agent, both Hit@1 and Recall@10 achieve their highest values on GPT-4.1 (25.93\% and 42.70\%, respectively).

\textbf{Analysis.} The results demonstrate that \tool consistently enhances issue localization across all three base models and baselines. At the file level, improvements are observed in all Hit@1 and Recall@10 values, ranging from 4.90\% to 166.19\% for Hit@1 and 6.15\% to 167.90\% for Recall@10. Similarly, at the function level, Hit@1 gains span 2.76\% to 504.35\% and Recall@10 ranges from 2.83\% to 376.13\%. These consistent gains indicate that \tool is effective regardless of the underlying LLM. The observed improvements across models of varying capabilities can be attributed to the clear and structured representation of concerns in \tool. Each concern contains semantically coherent functionalities, providing concise and high-quality context that can be readily interpreted by different models. This structured information allows smaller models to leverage the relevant repository knowledge effectively, while enhancing their issue localization~performance.


\summary{\tool consistently enhances issue localization across all three base models and baselines. For file-level localization, relative improvements range from 4.90\% to 166.19\% for Hit@1 and 6.15\% to 167.90\% for Recall@10. For function-level localization, Hit@1 gains span 2.76\% to 504.35\%, while Recall@10 improvements range from 2.83\% to 376.13\%.}

\subsection{RQ3: Ablation Study}

To investigate the necessity of incorporating concerns in \tool, we conduct an ablation study on three baselines. We use the same Hit@1 and Recall@10 as in \textbf{RQ2}. 

\textbf{Setup.} We incorporate three comparing versions with each baseline: 

\begin{itemize}[leftmargin=*]

    \item \textit{Ablating Term Explanation (w/o Exp)} does not rely on noun terms or concerns and instead directly leverages an LLM to generate summaries for all functions in the repository. The top $n$ functions are then selected based on the embedding similarity between their summaries and the issue.
    
    \item \textit{Ablating Concern Clustering (w/o Con)} utilizes \tool's Conceptual Term Extraction and Enrichment to construct a knowledge base, from which Issue-Specific Term Retrieval extracts relevant \textit{term functionalities}. Instead of applying Conceptual Concern Clustering and Candidate Concern Ranking, it directly computes the cosine similarity between the issue title and the retrieved \textit{term functionalities}. The top $n$ term functionalities are then selected according to their embedding similarity with the given issue.

    \item \textit{\tool} represents the original version, in which the top 10 concerns are selected.

    
\end{itemize}

\begin{table}
    \centering
    \footnotesize
    \renewcommand{\arraystretch}{0.9}
    \setlength{\tabcolsep}{2pt}
    \vspace{-3mm}
    \caption{Ablation Result}\label{tab:rq3}
    \begin{tabular}{lccccc}   
    \toprule
    \multirow{2}{*}{\textbf{Method}} & \multicolumn{2}{c}{\textbf{File Level}} & & \multicolumn{2}{c}{\textbf{Function Level}} \\ 
    \cmidrule{2-3} \cmidrule{5-6}
    & Hit@1 & Recall@10 & & Hit@1 & Recall@10 \\ 
    \midrule
    AgentLess & 15.28\%  &  25.33\% & &  10.19\% & 20.42\% \\ 
    \quad w/o Exp  & 28.70\%  & 38.52\% & & 19.91\% & 30.14\% \\ 
    \quad w/o Con  & 28.24\%  & 39.43\% & & 20.83\% & 33.09\% \\ 
    \quad \tool  & \textbf{\underline{29.17\%}} & \textbf{\underline{41.03\%}} & &
    \textbf{\underline{23.15\%}} &
    \textbf{\underline{34.95\%}} \\ 
    \hline
    
    OpenHands & 18.98\%  &  33.76\% & &  9.72\% & 17.72\% \\ 
    \quad w/o Exp  & 16.67\%  & 26.84\% & & 11.57\% & 16.45\% \\ 
    \quad w/o Con  & 14.81\%  & 29.68\% & & 10.19\% & 22.56\% \\ 
    \quad \tool  & \textbf{\underline{19.91\%}} & \textbf{\underline{40.89\%}} & &
    \textbf{\underline{14.35\%}} &
    \textbf{\underline{29.24\%}} \\ 
    \hline
    
    mini-SWE-agent & 24.07\%  &  44.21\% & &  16.67\% & 27.01\% \\ 
    \quad w/o Exp  & 21.76\%  & 33.87\% & & 12.96\% & 22.21\% \\ 
    \quad w/o Con  & 21.76\%  & 40.78\% & & 15.28\% & 28.83\% \\ 
    \quad \tool  & \textbf{\underline{25.46\%}} & \textbf{\underline{46.93\%}} & &
    \textbf{\underline{17.13\%}} &
    \textbf{\underline{36.32\%}} \\ 

    \bottomrule
    \end{tabular}
    
\end{table}

Based on our empirical findings that, on average, one concern contains 3.91 term functionalities in the concern-based approach, we scale the number of \textit{term functionalities} in the w/o Exp and w/o Con settings to align with \tool, which has approximately 39.1 term functionalities. Accordingly, we set $n=40$ for both w/o Exp and w/o Con. Additionally, to further validate the quality of the concerns in \tool, we compare the average \textit{Hit4File} and \textit{Hit4Func} scores over all files and functions. Here, \textit{Hit4File} and \textit{Hit4Func} indicate the Hit metric computed over all files or functions included in the three variants, respectively.

\textbf{Results.} Table \ref{tab:rq3} summarize our evaluation results for file-level and function-level issue localization. Overall, incorporating \tool with the top 10 concerns consistently improves localization performance across all three baselines. Specifically, at the function level for AgentLess, \tool achieves a Hit@1 of 23.15\%, outperforming \textit{w/o Exp} (19.91\%) and \textit{w/o Con} (20.83\%) by 3.24\% and 2.32\%, respectively. For OpenHands at the file level, \tool attains a Recall@10 of 40.89\%, surpassing \textit{w/o Exp} (26.84\%) and \textit{w/o Con} (29.68\%) by 14.05\% and 11.21\%. For mini-SWE-agent at the function level, \tool achieves a Recall@10 of 36.32\%, higher than \textit{w/o Exp} (22.21\%) and \textit{w/o Con} (28.83\%).

\begin{table}
    \centering
    \footnotesize
    \renewcommand{\arraystretch}{0.9}
    \setlength{\tabcolsep}{2pt}
    \caption{Hit Score in Different Context}\label{tab:rq3:compare}
    \begin{tabular}{lccc}   
    \toprule
    Metric & w/o Exp & w/o Con & \tool \\
    \midrule
    Hit4File & 0.37 & 0.53 & \textbf{\underline{0.62}} \\
    Hit4Func & 0.25 & 0.45 & \textbf{\underline{0.50}} \\
    \bottomrule
    \end{tabular}
\end{table}

Table \ref{tab:rq3:compare} demonstrates the quality of the concerns generated by \tool, regarding Hit4File and Hit4Func.
The results indicate that using \textit{Concerns} consistently achieves the highest Hit values at both file and function levels. Specifically, the Hit4File of \textit{Concerns} reaches 0.62, outperforming \textit{w/o Exp} (0.37) and \textit{w/o Con} (0.53) by 0.25 and 0.09, respectively. Similarly, the Hit4Func of \textit{Concerns} is 0.50, exceeding \textit{w/o Exp} (0.25) and \textit{w/o Con} (0.45) by 0.25 and 0.05, respectively. 

\textbf{Analysis.} Overall, leveraging high-level concerns in \tool consistently enhances localization performance across different baselines, yielding clear improvements at both file and function levels. Furthermore, our comparison of concern quality indicates that the contextual information captured by \textit{Concerns} is both more relevant and of higher quality, which accounts for the observed performance gains. Notably, \textit{w/o Con} outperforms \textit{w/o Exp}, suggesting that focusing on term-centric functionalities within functions is more effective than relying solely on general function summaries, as it captures term-specific semantics and helps mitigate concern tangling.

\summary{Incorporating \tool with noun terms and concerns consistently enhances localization performance. On average, \tool outperforms \textit{w/o Exp} by relative improvements of 2.47\% and 9.87\% in Hit@1 and Recall@10 for file-level localization, and 3.40\% and 10.57\% for function-level localization, respectively. Similarly, \tool outperforms \textit{w/o Con} by relative improvements of 3.24\% and 6.32\% in Hit@1 and Recall@10 for file-level localization, and 2.78\% and 5.34\% for function-level localization, respectively. Moreover, integrating concerns proves particularly effective in guiding the localization of both files and functions.}

\subsection{RQ4: Reliability of Concerns}
We evaluate the quality of the generated concerns through empirical human-judged metrics.

\textbf{Setup.} We randomly selected 30 issues and chose one concern from the top 10 for each, resulting in 30 concern-issue pairs. Each concern includes a descriptive summary and associated term functionalities, which are functional semantics from functions with specific noun terms clustered into the concern. We invite 10 developers with over three years of programming experience to evaluate the concerns. To ensure fairness, none had prior involvement in this work. We evaluate the quality of the generated concerns using three metrics:
\begin{itemize}[leftmargin=*]
    \item \textbf{Correctness}: The generated concern is clustered correctly, with every term functionality being relevant to its overall theme.
    \item \textbf{Completeness}: The concern summary fully captures the associated term functionalities.
    \item \textbf{Conciseness}: The concern summary contains no irrelevant or redundant information.
\end{itemize}

\begin{figure}
    \centering
    \vspace{-3mm}
    \includegraphics[width=\linewidth,keepaspectratio]{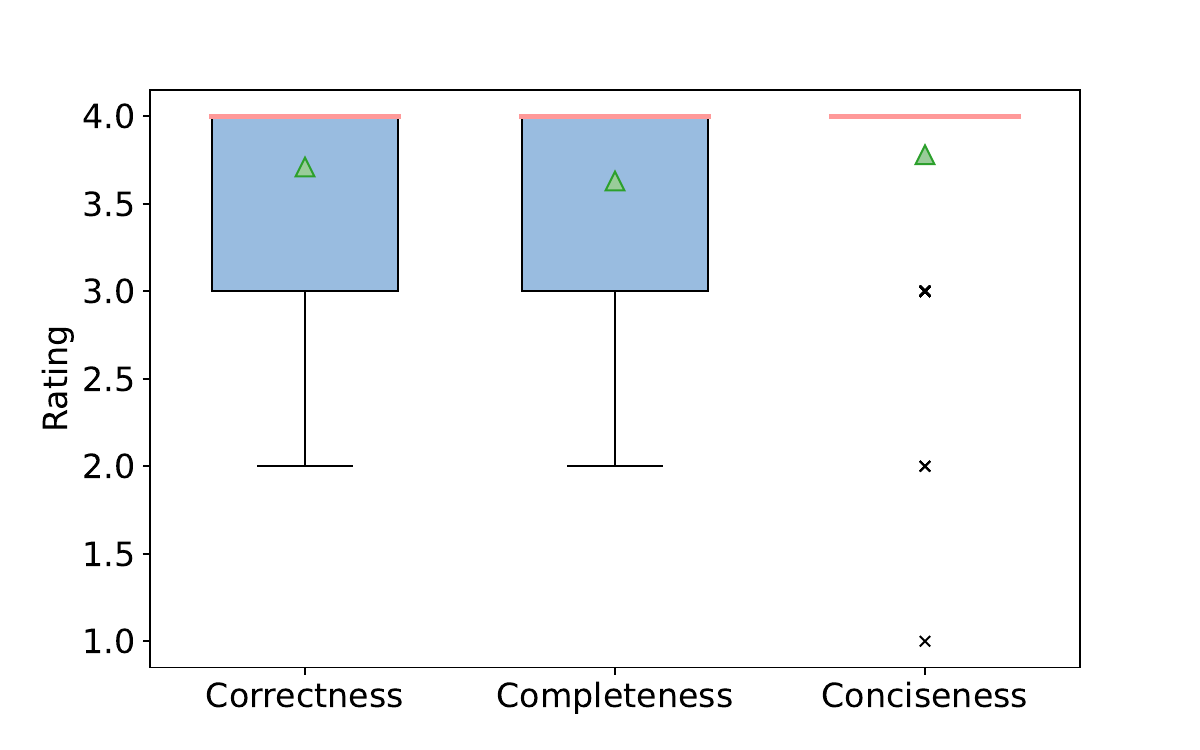}
    \captionof{figure}[Ratings of Correctness, Completeness, and Conciseness.]{Ratings of Correctness, Completeness, and Conciseness.} 
    \label{fig:rq4:ratings}
\end{figure}

\noindent Participants rate each concern on a 4-point Likert scale, where higher scores indicate better quality. All statements are phrased consistently in a positive form. After completing the ratings, participants are asked to clarify their judgments for any low scores (1 or 2).

\textbf{Results.} The results of the ratings for the quality of the generated concerns are shown in Figure~\ref{fig:rq4:ratings}. The ratings for the quality of the generated concerns demonstrate \tool's effectiveness across all three evaluation metrics. For correctness, 73.3\% of concerns were rated 4 (agree), 24.3\% rated 3 (somewhat agree), 2.3\% rated 2 (somewhat disagree), and none rated 1 (disagree), yielding an average score of 3.71. For completeness, \tool achieved an average score of 3.63, with 97.6\% of concerns receiving high scores (3 or 4) and the remainder rated 2. For conciseness, 79.7\% of concerns were rated 4, 19.0\% rated 3, 1.0\% rated 2, and only 0.3\% rated 1.

\textbf{Analysis. } The high correctness scores indicate that term functionalities were well-clustered and closely aligned with each concern’s theme. The completeness results show that the generated concerns preserve semantic integrity and include critical term functionalities, providing a comprehensive view of the repository knowledge. The conciseness ratings suggest that \tool presents information clearly and efficiently while maintaining semantic integrity, enhancing repository comprehension. Overall, these findings demonstrate that \tool consistently produces concerns that are correct, comprehensive, and concise.


\summary{97.6\%, 97.6\%, and 98.7\% of the sampled concerns were rated as correct, complete, and concise, respectively. The results indicate that \tool produces high-quality repository knowledge, facilitating developers with a clearer and more structured understanding.}

\section{Discussion}\label{sec-discuss}

\subsection{Limitations}

\textbf{Lack of Finer-grained Program Analysis.} First, although \tool considers structural relationships such as function calls during concern clustering, it does not leverage more advanced program analysis techniques, such as program slicing, control-flow or data-flow analysis, and dynamic execution tracing. As a result, it may miss subtle dependencies or execution paths that are critical for understanding complex code behavior, potentially limiting the precision of the constructed concerns in capturing all relevant functionalities for an issue.

\textbf{Lack of Advanced Integration Techniques.} Second, \tool integrates concerns into existing LLM-based localization methods via prompt modifications rather than employing a dedicated, concern-aware localization agent. This design choice, while minimally intrusive, may prevent the localization models from fully exploiting the semantic structure and relationships provided by the concerns. Consequently, there may be additional gains in localization performance achievable with tighter integration or agent-level reasoning that explicitly models concern semantics during the search for faulty code.

As future work, we plan to incorporate advanced program analysis techniques and design specialized agents that exploit deep research mechanisms to maximize the utility of concerns.

\subsection{Threats to Validity}
\textbf{Internal Threats.} The threats to the internal validity of our study mainly lie in the randomness of LLM outputs and the subjectiveness of human evaluation. To address these, we fix random seeds and set the model temperature to 0 where applicable. For human evaluation, we involve multiple annotators, provide detailed guidelines, and measure inter-annotator agreement to ensure consistency and reduce bias. 

\textbf{External Threats.} The threats to the external validity arise from the single benchmark and specific LLMs we use, which may limit the generality of our findings. To mitigate these, we evaluate on all 216 issues of varying difficulty in the benchmark, and as future work, we plan to extend \tool to additional benchmarks and more diverse models.

\section{Related Work}\label{sec-related}

\textbf{Issue Localization.}
Early issue localization  techniques primarily relied on information retrieval (IR) approaches~\cite{DBLP:conf/icse/ZhouZL12, DBLP:conf/kbse/SahaLKP13, DBLP:conf/icsm/SahaLKP14, DBLP:journals/smr/SismanAK17}, which rank source files or methods by measuring textual similarity between bug reports and code artifacts. Additionally, some methods integrate dynamic runtime information to improve localization. For example, Spectrum-Based Fault Localization (SBFL)~\cite{DBLP:conf/prdc/AbreuZG06, DBLP:journals/tr/WongDGL14, DBLP:journals/corr/SouzaCK16} statistically associates test outcomes with executed code elements, ranking them by suspiciousness derived from passing and failing test coverage. Mutation-Based Fault Localization (MBFL)~\cite{DBLP:journals/stvr/PapadakisT15} instead introduces artificial faults (mutants) into the program and evaluates suspiciousness based on how effectively test cases expose these mutants. While effective in controlled settings, these dynamic approaches are limited in practice: they cannot exploit the rich context in bug reports, and they depend heavily on the availability of high-quality test cases—an assumption often unmet in large-scale repositories.

More recently, LLMs have driven significant progress in issue localization. Some studies directly prompt LLMs to identify defective code elements~\cite{DBLP:journals/corr/abs-2308-15276}, while others fine-tune models to predict bug locations from the joint semantics of code and bug reports~\cite{DBLP:conf/icse/CiborowskaD22, DBLP:journals/access/MohsenHWMM23, DBLP:conf/icse/YangGMH24}. Beyond single-function prediction, recent work has begun addressing repository-level localization—identifying relevant code segments across large projects~\cite{DBLP:journals/corr/abs-2407-01489, DBLP:journals/corr/abs-2405-15793, DBLP:conf/iclr/0001LSXTZPSLSTL25, DBLP:journals/corr/abs-2502-15292, DBLP:conf/acl/ChenTDW0JPC025, DBLP:journals/corr/abs-2502-00350, DBLP:journals/corr/abs-2503-22424, DBLP:journals/corr/abs-2505-07849, DBLP:journals/corr/abs-2505-19489, DBLP:conf/icse/BatoleON0R25}. These approaches are often built on agent-based systems that treat bug localization as a planning and search problem.
While promising, these LLM- and agent-based approaches still face challenges in interpreting mixed concerns and scattered responsibilities in large codebases. Without explicit modeling of conceptual responsibilities, they are prone to drifting or misinterpreting the distributed nature of fault-relevant logic. 
This highlights the need for methods that explicitly abstract and align domain concepts with their diverse and scattered implementations—a gap our approach aims to fill.

\textbf{Retrieval-Augmented Generation.}
Retrieval-Augmented Generation (RAG) has emerged as a prominent technique in the LLM literature for enhancing generation quality and factual accuracy. It extends an LLM’s context by incorporating documents or knowledge retrieved from an external corpus. The framework, first formalized by Lewis et al.~\cite{DBLP:conf/nips/LewisPPPKGKLYR020} in 2020, typically involves a two-stage process: a retriever first fetches relevant documents, and then a generator conditions its output on both the original query and the retrieved documents. This interaction improves output accuracy by grounding generation in explicit evidence, rather than relying solely on the language model’s internal knowledge.

RAG has been widely applied to software engineering tasks, including code generation~\cite{DBLP:conf/iclr/Zhou0XJN23, DBLP:conf/naacl/WangAYXXNF25}, code completion~\cite{DBLP:conf/acl/LuDHGHS22, DBLP:conf/emnlp/ZhangCZKLZMLC23}, vulnerability detection~\cite{DBLP:journals/corr/abs-2406-11147, DBLP:journals/corr/abs-2407-14838}, and program repair~\cite{DBLP:conf/sigsoft/Wang0JH23}. In this paper, we likewise employ RAG to enhance agent-based understanding of a target project. Unlike prior work that primarily retrieves documents or raw code snippets, our approach constructs and retrieves finer-grained conceptual knowledge units, referred to as Concerns. By grounding retrieval in concept-level abstractions, our method provides more accurate and robust evidence to support downstream LLM reasoning for issue localization.


\section{Conclusion}\label{sec-conclusion}
We presented \tool, an approach that enhances issue localization by abstracting fine-grained functionalities into high-level conceptual concerns. This addresses challenges such as concern tangling and concern scattering and provides LLMs with structured guidance for efficient and accurate code exploration. Experiments show that \tool consistently improves state-of-the-art localization tools across multiple models, while ablation studies and manual evaluation confirm the reliability of the constructed concerns. These results demonstrate the effectiveness of leveraging conceptual knowledge for issue localization in large-scale software systems.

\bibliographystyle{ACM-Reference-Format}
\bibliography{main}

\end{document}